\documentclass{article}
\usepackage{ijcai16}

\usepackage{times}
\usepackage{amsmath,amssymb}
\usepackage{graphicx} 
\usepackage{subfig}
\usepackage{multirow}
\usepackage{hyperref}
\usepackage{changepage}
\usepackage{adjustbox}
\usepackage[font=small]{caption}


\makeatletter
\renewcommand\@biblabel[1]{} % No brackets for the references
\renewenvironment{thebibliography}[1]
     {\section*{\refname}%
      \@mkboth{\MakeUppercase\refname}{\MakeUppercase\refname}%
      \list{\@biblabel{\@arabic\c@enumiv}}%
           {\settowidth\labelwidth{\@biblabel{}}%
            \leftmargin\labelwidth
            \advance\leftmargin10pt% change 20 pt according to your needs
            \advance\leftmargin\labelsep
            \setlength\itemindent{-10pt}% change using the inverse of the length used before
            \@openbib@code
            \usecounter{enumiv}%
            \let\p@enumiv\@empty
            \renewcommand\theenumiv{\@arabic\c@enumiv}}%
      \sloppy
      \clubpenalty4000
      \@clubpenalty \clubpenalty
      \widowpenalty4000%
      \sfcode`\.\@m}
     {\def\@noitemerr
       {\@latex@warning{Empty `thebibliography' environment}}%
      \endlist}
\renewcommand\newblock{\hskip .11em\@plus.33em\@minus.07em}
\makeatother

\usepackage{latexsym} 

\title{Face valuing: Training user interfaces with facial expressions\\ and reinforcement learning} %\thanks{These match the formatting instructions of IJCAI-07. The support of IJCAI, Inc. is acknowledged.}
\author{Vivek Veeriah, Patrick M. Pilarski, Richard S. Sutton \\ 
Reinforcement Learning \& Artificial Intelligence Lab \\
Department of Computing Science, University of Alberta  \\
\texttt{\{vivekveeriah, pilarski, rsutton\}@ualberta.ca}}

\begin{document}

\maketitle

\begin{abstract}
An important application of interactive machine learning is extending or amplifying the cognitive and physical capabilities of a human. To accomplish this, machines need to learn about their human users' intentions and adapt to their preferences. In most current research, a user has conveyed preferences to a machine using explicit corrective or instructive feedback; explicit feedback imposes a cognitive load on the user and is expensive in terms of human effort. The primary objective of the current work is to demonstrate that a learning agent can reduce the amount of explicit feedback required for adapting to the user\rq s preferences pertaining to a task by learning to perceive a value of its behavior from the human user, particularly from the user\rq s facial expressions---we call this {\em face valuing}.  We empirically evaluate face valuing on a grip selection task. Our preliminary results suggest that an agent can quickly adapt to a user\rq s changing preferences with minimal explicit feedback by learning a value function that maps facial features extracted from a camera image to expected future reward. We believe that an agent learning to perceive a value from the body language of its human user is complementary to existing interactive machine learning approaches and will help in creating successful human-machine interactive applications.
\end{abstract}
%%(in contrast to most existing work, wherein facial features are used directly as control signals). 

\section{Introduction}

% Humans can effortlessly interact with fellow humans during their day-to-day interactions. However, when it comes to humans and machines, it is rather difficult to establish a medium of intuitive communication. Despite this, it is widely believed that humans and machines can work in unison to solve hard challenges facing humanity. The field of research that addresses this compelling problem is called human-machine interaction.

One important objective of human-machine interaction is to augment existing human capabilities, which requires machines and their human users to closely collaborate and form a productive partnership. To achieve this, it is crucial for the machines to learn interactively from their users, specifically their intents and preferences. In current research trends, the user\rq s preferences are conveyed via explicit instructions (Kuhlmann et al., 2004) or expensive corrective feedback (Knox \& Stone, 2015)---which can be in the form of predefined words or sentences, push buttons, mouse clicks etc. In many real-world, ongoing scenarios, these methods of feedback impose a cognitive load on human users. Moreover, in complex domains like prosthetic limbs, it is demanding for the user to provide these kinds of explicit feedback. It is important to have an alternative approach that is both scalable and would allow the machines to learn their human users\rq intents and preferences via ongoing interactions.

In this paper, we explore the idea that a reinforcement learning agent can learn a value function that relates a user's body language, specifically from the user's facial expressions, to expectations of future reward. The agent can use this value function to adapt its actions to a user's preferences quickly with minimal explicit feedback. This approach is analogous to an agent learning to understand the body language of its human user. It could also be imagined as building a form of communicative capital between a human user and a learning agent (c.f., Pilarski et al., 2015). Learning from interactions with a human user tend to be continual; reinforcement learning methods are therefore naturally suited for this purpose.

To the best of our knowledge, our system is the first to learn a value function in real-time for a user's body language, specifically a value function that relates future reward to the facial features of the user. Additionally, this work is the first example of how such a value function can be used to guide the learning process of an agent interacting with a human user. Importantly, our approach does not utilize explicit reward channels, for example those discussed by Thomaz \& Breazeal (2012) and by Knox et al. (2009).  %Moreover, significantly differs from the various other perspectives adopted by the interactive machine learning community for developing user interactive agents.
%Through our preliminary experiments, our approach seems to be a promising framework towards intuitive human-machine interaction. 
As it operates in real time, we believe that our approach is well suited for realistic human-machine interaction tasks and complements existing interactive machine learning approaches. Learning a language between an agent and its user in the form of value functions represents a new and powerful class of human-machine interaction technologies, and we expect the approaches discussed in this work will have broad applicability in many different real-world domains.

%Learning to adapt to the changing preferences of a human user is an interesting problem that arises in many other human-machine interactive applications as well. To this end, we propose that by enabling the agent to learn to perceive a value of its behavior from its human user, the the agent could quickly adapt its behavior with regard to the user's preference, without receiving explicit user generated rewards. From our experiments, it was observed that the agent had learned to perceive a value from the human user's face with respect to a task. However, we believe our approach to be general i.e. the agent could perceive a value of its behavior from any relevant signal source and not just a human's face.

\section{Related Methods}

Significant research effort has been directed toward creating successful human-robot partnerships (e.g., as summarized in Knox \& Stone, 2015; Mead et al., 2013; Breazeal et al., 2012; Pilarski \& Sutton, 2012; Edwards et al., 2015). A natural approach is for an agent to learn from ongoing interactions with a human user via {\em human-delivered  rewards}. Research by Thomaz \& Breazeal (2008), Knox \& Stone (2009), Breazeal et al. (2012), Loftin et al. (2015), and Peng et al. (2016) adopts this perspective, and it has been extensively studied in recent work by Knox \& Stone (2015). In the TAMER approach of Knox and Stone (2015), a system was able to learn a predictive model of a human user\rq s shaping rewards, such that the model could be used to successfully train an agent even in the presence of human-related delays and inconsistencies. As a potential drawback of learning a reward model, when the user needs to modify the agent\rq s behavior, the model would have to be changed (e.g., via additional shaping rewards from the user). We desire a method for delivering feedback that does not require a large number of costly interactions from the human, and that transfers well to new or changed situations without the need for retraining.%Moreover, human-generated rewards are impractical to obtain in many realistic scenarios.

Another interesting approach to the interactive instruction of a machine learner involves a human teaching a robot to perform a task through demonstrations, a process aptly named as {\em learning from demonstration}. This approach can also be called programming by demonstration. There are numerous works exploring learning from demonstration (e.g., Koenig \& Mataric, 2012; Schulman et al., 2013; Alizadeh et al., 2014). One noted downside is that this form of learning is reported to be at times a tiring experience for a human user. Many approaches are also limited in their ability to scale up to a full range of real-world tasks (e.g., it is impossible to tractably provide demonstrations or instructions covering all possible situations).

A key difference between many existing methods and our approach is that we are concerned with designing a general, scalable approach that would allow an agent to adapt its behavior to a user\rq s changing preferences with minimal explicit human-generated feedback. This is in contrast to approaches that seek to use body language like facial features as an input channel to directly control a robot or other machine's operation (e.g.,  Breazeal (1998) and Liu \& Picard (2003)). As a significant contribution of the present work, we describe the use of facial features not as a channel of control but as a means of valuation. To this end, we propose to learn a value function that is grounded in the user\rq s body language, independent of the features of a task, and to use this value function to help influence an agent's real-time decision making in a way that spans multiple tasks and settings of use.

\section{Reinforcement Learning}

In a reinforcement learning setting (Sutton \& Barto, 1998), a learning agent interacts with an unknown environment in order to achieve a goal. In this setting, the goal is to maximize the cumulative reward accumulated by adapting its behavior within the environment. 

Markov Decision Processes (MDPs) are the mathematical notations used for formalizing a reinforcement learning problem. An MDP consists of a tuple \big \langle $ \mathcal{S}, \mathcal{A}, p, r, \gamma $ \big \rangle, consisting of $ \mathcal{S} $, set of all states; $ \mathcal{A} $, set of all actions; $ p(s^{\prime} | s, a) $, a transition probability function, which gives the probability of transition to state $ s^{\prime} \in \mathcal{S} $ at the next time-step given for the current state $ s \in \mathcal{S} $ and action $ a \in \mathcal{A} $; $ r(s, a, s^{\prime}) $, the reward function that gives the expected reward for a transition from state $ s \in \mathcal{S} $ to $ s^{\prime} \in \mathcal{S} $ by taking action $ a \in \mathcal{A} $; $ \gamma $ is the discount factor, that specifies the relative importance between immediate and long term rewards. In episodic problems, the MDP can be viewed as having special states called \textit{terminal states}, which terminate an episode. Such states ease the mathematical notations as they could be viewed as a single state with single action that results in a reward of 0 and transition to itself. The return at a time instance \textit{t} is defined as the discounted sum of rewards after time \textit{t}:

\begin{align*} % requires amsmath; align* for no eq. number
   G_{t} & = R_{t + 1} + \gamma R_{t + 2} + \gamma^{2} R_{t + 3} + \cdots \\
   G_{t} & =  \sum_{i = 1}^{\infty} \gamma^{ i - 1} R_{t + i} 
\end{align*}

where $ R_{t + 1} $ denotes the reward received after taking an action $ A_{t} $ in state $ S_{t} $.

Actions are taken at discrete time steps $ t = 0, 1, 2, \cdots $ according to a \textit{policy} $ \pi : \mathcal{S} \times \mathcal{A} \rightarrow [0, 1] $ which defines a selection probability for each action conditioned on the state. Each policy $ \pi $ has a corresponding state-value function $ v_{\pi}(s) $, that maps each state $ s \in \mathcal{S} $ to the expected return $ G_{t} $ from that state by following the policy $ \pi $, 

\begin{align*} % requires amsmath; align* for no eq. number
   v_{\pi}(s) = \mathbf{E} \big \{ G_{t} | S_{t} = s, \pi \big \}
\end{align*}

The state-value functions are significant when the given task requires prediction. On the contrary, if the given task requires control, then it is important to use the action-value functions $ q_{\pi}(s, a) $ which gives the expected return $ G_{t} $ by taking an action $ a $ from state $ s $ and then following the policy $ \pi $: 

\begin{align*} % requires amsmath; align* for no eq. number
  q_{\pi}(s, a) = \mathbf{E} \big \{ G_{t} | S_{t} = s, A_{t} = a, \pi \big \}
\end{align*}

\section{Grip Selection Task}
 
\begin{figure*}[ht!]
%\begin{adjustwidth}{-0.75cm}{}
\centering
\begin{tabular}{ccc}
\subfloat[]{\includegraphics[keepaspectratio,width=9.5cm,height=5.5cm]{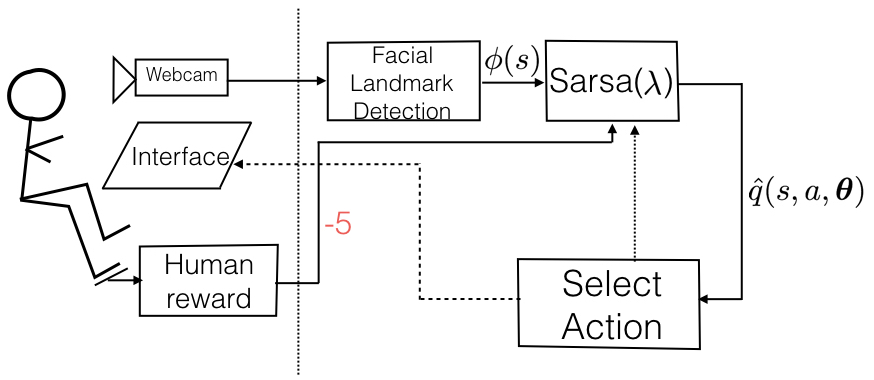}} & 
\subfloat[]{\includegraphics[keepaspectratio,width=7.5cm,height=5.5cm]{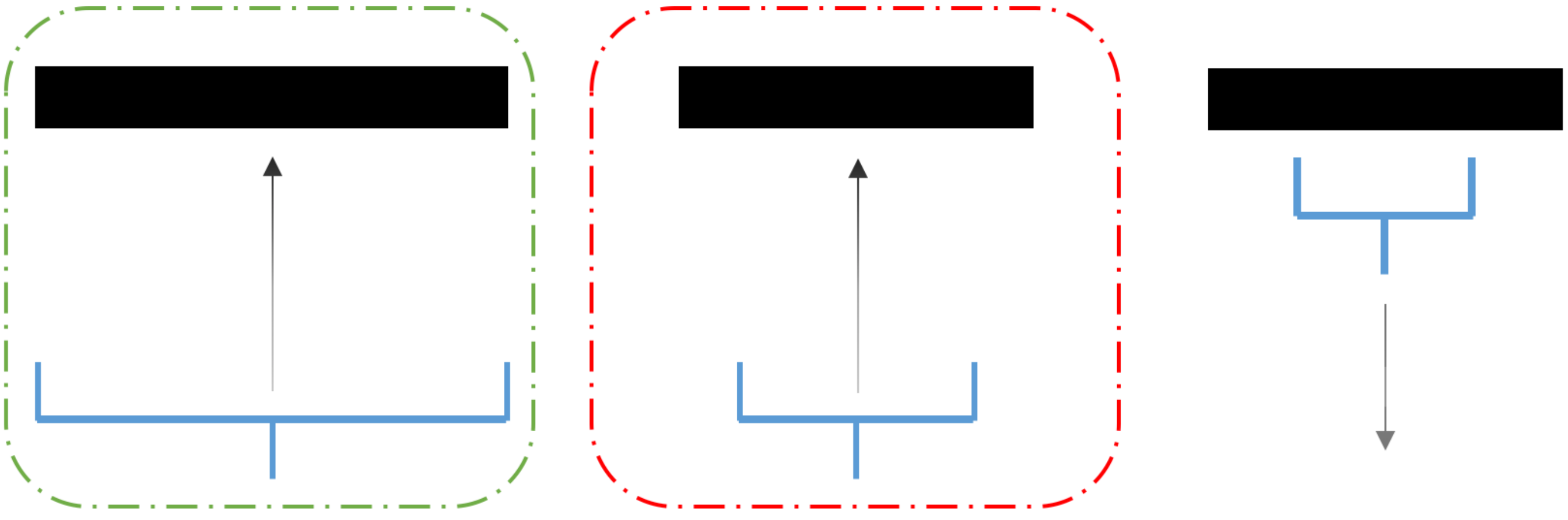}} 
\end{tabular}
\caption{(a) Overview of the face valuing agent: the user observes the simulated task and can provide a negative reward to the SARSA agent during the task. The agent can observe user\rq s face via a webcam, and learns to behave based on the user\rq s preferences. (b) Grip selection task: Solid colored box denotes the object; the thin blue lined object denotes the grip selected by an agent. The initial position of the agent is called ``grip-changing station". The agent has to learn to select, based on the user\rq s preference, an appropriate grip --- the grip\rq s width needs to be equal to the object\rq s width. The figure on the left denotes one correct combination while the figure on the center denotes an incorrect grip for the given object. The figure on the right denotes a scenario where the agent is forced to return to its initial position, as the user has pushed the reward button. To complete an episode, the agent has to select one of the correct grips; move forward and grasp the object.}
 \label{fig:tasksetup}
%\end{adjustwidth}
\end{figure*}
 
To evaluate the face valuing approach, we introduce a grip selection task that was inspired by a natural problem in a prosthetic arm domain where the agent needs to select an appropriate grip pattern for grasping a given object. The task consists of a set of \textit{n} grips and \textit{m} objects. Depending on the experiment, there could be many possible grips for a given object, with the correct grip being defined according to the user\rq s preference. The task could also consist of uncountable number of objects, making grip selection with pure trial-and-error a non-trivial problem.% it is not viable for the agent to select a correct grip through trial and error

This task was formulated as an undiscounted episodic MDP with a reward of 0 for every time step, and with 0 reward for completing an episode. At the beginning of each episode, a single object was randomly picked from a large set of objects and the agent was tasked with choosing a grip from a limited set of grips; once the agent selected a grip it needed to move a fixed number of steps towards the object to finish the grasping motion, thereby completing an episode. A human user provided reward to the agent by pressing a single button which delivered a reward of -5 for the corresponding time step. Moreover, pressing the button forced the agent to return back to its initial position regardless of its current position. The experimental setup is shown in Fig. \ref{fig:tasksetup}.

%\begin{figure}
%\centering
%\includegraphics[keepaspectratio,width=6cm,height=5cm]{figures/gripsetup_1.pdf}
%\caption{Grip selection task: Solid colored box denotes the object; the thin blue lined object denotes the grip selected by an agent. Agent needs to select an appropriate grip - the grip\rq s width needs to be equal to the object\rq s width. The figure on the left denotes one correct combination while the figure on the right denotes an incorrect grip for the given object. To complete an episode, the agent has to select one of the correct grips; move forward and grasp the object.}
%\label{fig:tasksetup}
%\end{figure} 

As in real-world grasping tasks, a user could have personal preferences over which grip was suitable to grasp a given object. These preferences could change from episode to episode and from experiment to experiment. Further, these preferences were hidden from the learning agent and the only way the agent could infer this is from the changing facial expressions of the human user. Therefore, in this work, we asked the human user to be as expressive as possible so as to provide clear cues for the learning agent to begin forming its behavioral choices. 

\section{Experimental Setup}
For our experiments, two Sarsa($ \lambda $) (Rummery \& Niranjan, 1994; Sutton \& Barto, 1998) agents (one that uses face valuing and one without face valuing) are compared on the above described task. All the experimental results in this paper were performed by a well-trained user in a blind setting, i.e., the user did not know which of the two methods was currently being evaluated. The user provided the same form of rewards to both learning agents via their button pushes. The two main instructions we gave to the human user were 1) to express their pleasure or displeasure with the agent via any simple, repeatable, and minutely distinguishable facial expressions, and 2) to push the button whenever the learning agent was not behaving as per the user's expectations. 

%
%\subsection{Feature extraction from a human's face}
%
%For the objective motivated in this paper, we required simple and efficient face detection and feature extraction methods. For our experiments, we utilized a popular face detection algorithm originally designed by \cite{viola2004robust} and a facial landmark extraction algorithms based on the work of \cite{kazemi2014one}. Though the facial landmark detection algorithms were intended for aligning faces from multiple images, they are very well suited for our purpose. 
%
%The landmark detection algorithm detects the position of 68 key points from a frame containing a human's face. The points from each frame were normalized. Among these 68 points, only 23 points correspond to the positions of eye brows and mouth of a human's face. Presumably, we expect them to be important for classifying facial expressions and so, only these points were selected. Each of these 23 points, were tile-coded with $ 4 $ tilings and each tiling of size $ 10 \times 10 $ resulting in a feature vector of size 9200. 
%
%\begin{figure}
%\centering
%\includegraphics[keepaspectratio,width=4cm,height=4cm]{figures/featureextraction.png}
%\caption{Features extracted from face: facial landmarks corresponding to eye brows and mouth of a user\rq s face are the only features extracted from the face.These are marked as blue circles.}
%\label{fig:featureextraction}
%\end{figure}
%
%The features extracted from the human user's face could be better understood from the fig. \ref{fig:featureextraction}.

\subsection{State space}

\begin{table}[h!]
\centering
\begin{adjustbox}{width=0.5\textwidth}
\small
 \begin{tabular}{| c c |} 
 \hline
  Agent & State space \\
 \hline \hline
w/o face valuing & current grip and object ids + bias term \\
w/ face valuing & 23 feature points + bias term\\
 \hline
 \end{tabular}
 \end{adjustbox}
 \caption{State space of the agents compared in the experiments are displayed.}
 \label{tab:statespace}
\end{table}

The state spaces for both the agents are briefly described in Table \ref{tab:statespace}. The agent without face valuing has the id of the current grip and id of the current object in its state space along with a bias term. The id of the current grip chosen by the agent is one-hot encoded to form vector of length $n$. Similarly, the id of the object is also one-hot encoded to a vector of length $m$ and concatenated with the vector corresponding to the current grip. So, the entire state space for this method is of length $m + n + 1$ where $m$ is the total number of objects during the entire experiment and $n$ is the total number of grips available to the agent during an experiment.

For the face valuing agent, 68 key points from a frame containing a human\rq s face are detected through a popular facial landmark detection algorithm (Kazemi et al., 2014). These key points are simple two dimensional coordinates that denote the position of certain special locations of a human\rq s face. These points from each frame were normalized and 23 points that correspond to the positions of eye brows and mouth of a human's face were selected. Each of these 23 points, were tile-coded with $ 4 $ tilings and each tiling of size $ 10 \times 10 $ resulting in a feature vector of size 9200. These key points seems to produce sufficient variations between different facial expressions.

\begin{figure}[h!]
\centering
\includegraphics[keepaspectratio,width=5.5cm,height=6cm]{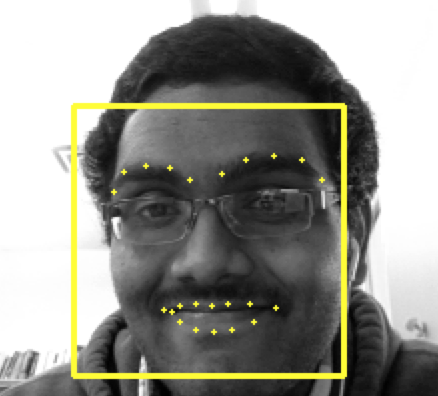}
\caption{Features extracted from face: facial landmarks corresponding to eye brows and mouth of a user\rq s face are the only features extracted from the face.These are marked as yellow circles.}
\label{fig:featureextraction}
\end{figure}

%The features extracted from the human user's face could be better understood from the fig. \ref{fig:featureextraction}.

\begin{figure*}[t]
\vspace{-1.5cm}
\begin{adjustwidth}{-1.0cm}{}
\centering
\begin{tabular}{ccc}
\subfloat[2 objects \& 2 grips]{\includegraphics[keepaspectratio,width=6cm,height=6cm]{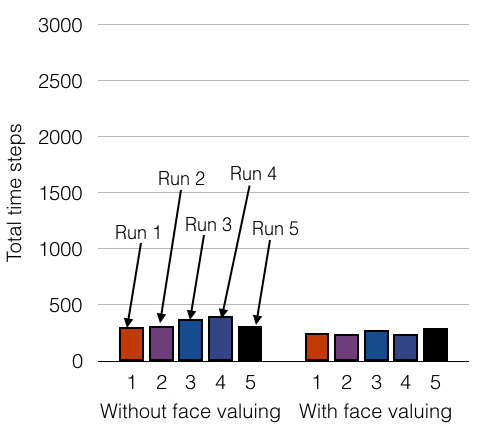}} & 
\subfloat[2 objects \& 4 grips]{\includegraphics[keepaspectratio,width=6cm,height=6cm]{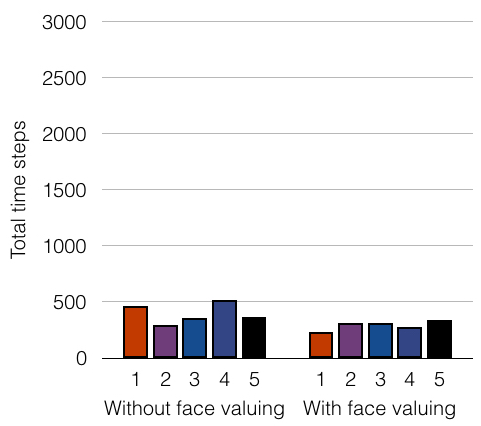}} & 
\subfloat[2 objects \& 8 grips]{\includegraphics[keepaspectratio,width=6cm,height=6cm]{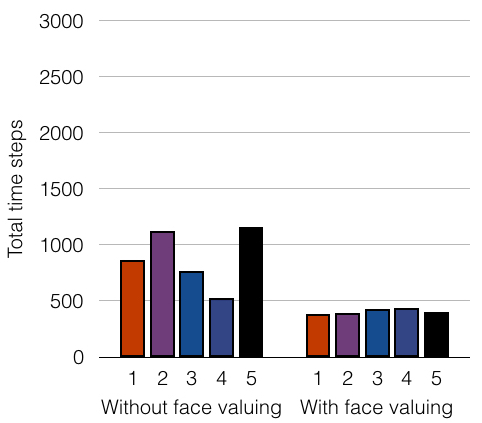}} \\
\subfloat[4 objects \& 2 grips]{\includegraphics[keepaspectratio,width=6cm,height=6cm]{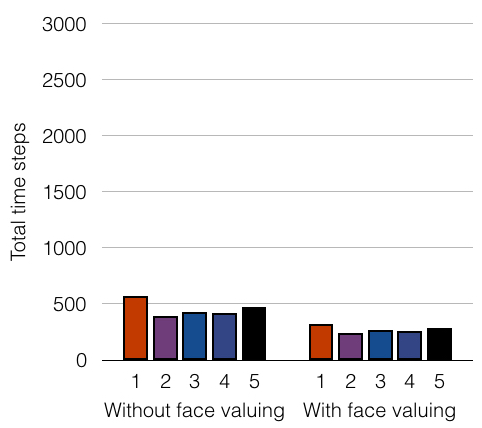}} &
\subfloat[4 objects \& 4 grips]{\includegraphics[keepaspectratio,width=6cm,height=6cm]{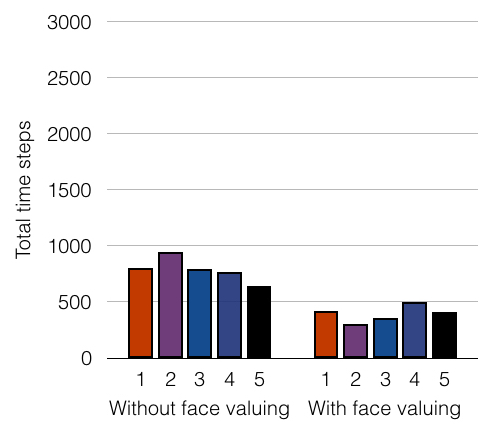}} &
\subfloat[4 objects \& 8 grips]{\includegraphics[keepaspectratio,width=6cm,height=6cm]{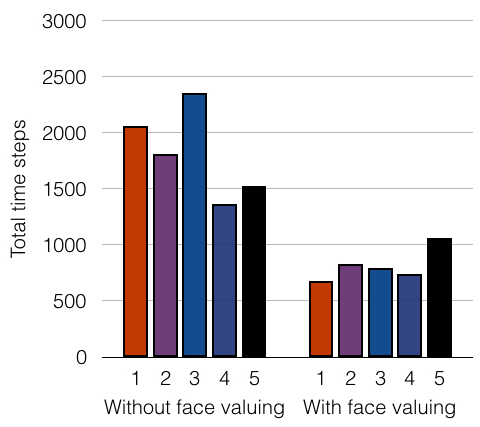}} \\
\subfloat[8 objects \& 2 grips]{\includegraphics[keepaspectratio,width=6cm,height=6cm]{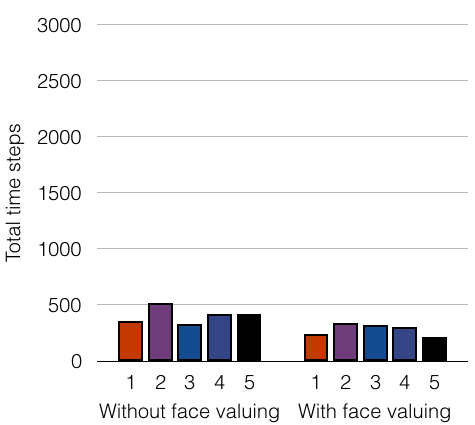}} &
\subfloat[8 objects \& 4 grips]{\includegraphics[keepaspectratio,width=6cm,height=6cm]{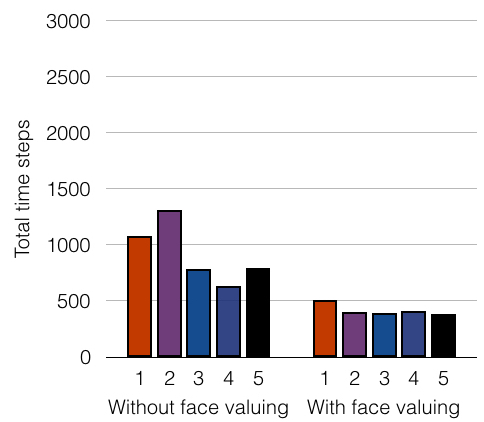}} &
\subfloat[8 objects \& 8 grips]{\includegraphics[keepaspectratio,width=6cm,height=6cm]{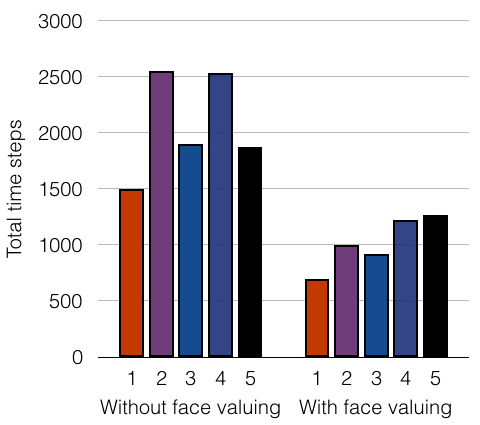}} \\
\subfloat[2 grips setting]{\includegraphics[keepaspectratio,width=6cm,height=6cm]{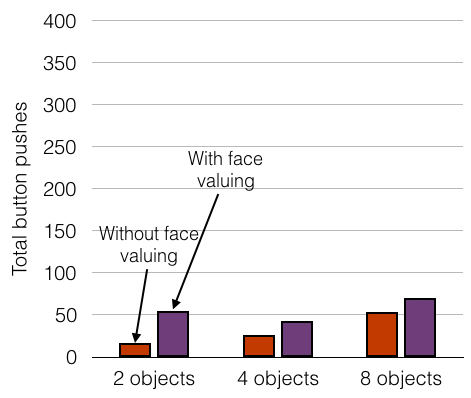}} &
\subfloat[4 grips setting]{\includegraphics[keepaspectratio,width=6cm,height=6cm]{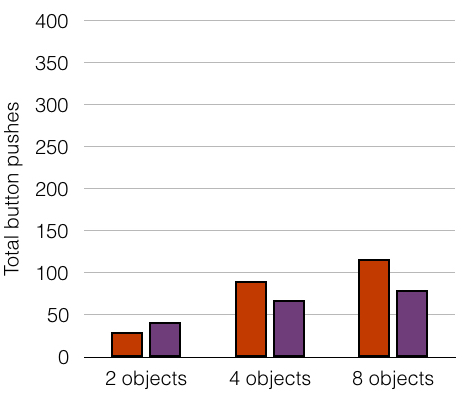}} & 
\subfloat[8 grips setting]{\includegraphics[keepaspectratio,width=6cm,height=6cm]{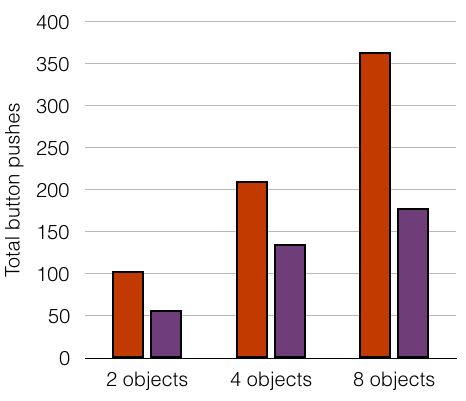}}
\end{tabular}
\caption{Total time steps taken for different grip and object settings: the face-valuing agent learned to adapt faster compared to a conventional agent in most settings. Moreover, in difficult settings where there are more number of appropriate object-grip combinations, the face-valuing agent required less human generated rewards for achieving this level of performance. Each plot was generated by averaging data obtained from 5 independent runs and each run consisted of 15 episodes.}
 \label{fig:learningcurves}
\end{adjustwidth}
\end{figure*}

\clearpage

\subsection{Action space}
The complete action space for both the agents consisted of $ \{  grip_{1}, grip_{2}, \cdots, grip_{n}, \uparrow, \downarrow \} $ actions where the first $ n $ actions implied in choosing that particular grip. The remaining two actions were the move one step forward towards the object and move one step back towards the grip-changing station. The actions available to the agent depended on its position relative to the object and the grip-changing station. When the agent was in the grip-changing station, the available actions were $ \{ grip_{1}, grip_{2}, \cdots, grip_{n}, \uparrow \} $ whereas when the agent had left the grip-changing station, only the $ \{ \uparrow, \downarrow \} $ actions were available. When the user pushed the reward button, the agent lost all its actions except $ \{ \downarrow \} $ until it reached the grip-changing station.

The agent observed the state space once every {\em one-tenth of a second} and had to take an action on every time step. The agent, however, had the freedom of choosing the same action for many consecutive time steps which allowed the human user to expressively respond to the learning agent. 

\section{Experiments}

\subsection{Experiment 1: Different object-grip settings}

The first experiment compared the two agents with multiple grip \& object settings. The plots of total time steps and total human generated rewards accumulated by the agents are shown in Fig. \ref{fig:learningcurves}. The plots (Fig. \ref{fig:learningcurves} (a), (b), $ \cdots $, (i)) represent the total time taken by a learning agent to complete a successful grasp across episodes. The plots (Fig. \ref{fig:learningcurves} (j), (k), (l)) display the total number of human generated rewards given to an agent to successfully adapt to user\rq s preferences. These graphs (Fig. \ref{fig:learningcurves}) were generated from the same user experiments conducted in a blind manner. A perfect agent would have no human generated reward in all these settings and would take only 11 time steps to complete an episode. 

From the plots (Fig. \ref{fig:learningcurves} (a), (b), $ \cdots $, (i)), it can be observed that the agent with face valuing quickly adapted with the user\rq s preferences in all the different object and grip settings. Also, from the plots (Fig. \ref{fig:learningcurves} (j), (k), (l)), the total number of human generated rewards for the face valuing agent was comparatively lower than the agent without face valuing in the difficult settings of this experiment.

During the initial phase of the experiment, the face valuing agent utilized the human generated reward to learn to perceive a value of its actions from the human user\rq s face. This learned value was leveraged to adapt the agent\rq s actions in later phases of the experiment. In simple experiment settings where there are fewer number of object-grip combinations, like the 2 grips experiment setup, the agent without face valuing could quickly learn the appropriate behavior from button pushes provided by the user and this resulted in a better performance compared to an agent with face valuing. However, in setups with large number of possible combinations of grips and objects, the agent without face valuing lost this advantage and failed to scale up. The face-valuing agent performed comparatively better in these scenarios as it learned to perceive a ``goodness" of its actions which guided the agent in choosing the correct action at a given time instance.

For the agent without face valuing, the user\rq s preferences could be communicated only through the reward channel. Naturally, this approach was more expensive in terms of the number of manual rewards compared to the face-valuing approach. On the other hand, the face-valuing approach utilized the user\rq s facial features, which conveyed the preferences over the grips---a simple approach observed in the user was to for them to have a neutral or a sad expression when the agent was not selecting the correct grip and to display a positive expression when the agent selected the correct grip. Interestingly, by learning a value function over the face, this agent learned to wait for an affirmative expression from the human user before moving forward to grasp the object. When there were no such expressions from the user, the agent switched from one grip to another until the user gave a ``go ahead" expression. %This sort of behavior in an agent would be desirable in the prosthetic limb domain. <NOT SURE THIS IS SUPPORTED YET>

\subsection{Experiment 2: Infinite objects and finite grips}

%\begin{figure*}
%\centering
%\begin{tabular}{cc}
%\subfloat[Infinite objects \& 5 grips]{\includegraphics[keepaspectratio,width=6cm,height=6cm]{figures_upd/infiniteobjects_1.eps}} &
%\subfloat[Learning curve]{\includegraphics[keepaspectratio,width=6cm,height=6cm]{figures_upd/infobj_learningcurve.eps}}
%\end{tabular}
%\caption{Plot of total time steps for limited grips and infinite objects setting: an experiment setting where a new object is introduced every episode and the agent needs to grasp this object from one of its grip. Each bar represents the total time steps taken by the agent to complete 15 episodes of this task. The plot was generated from data obtained from 5 independent runs. This is a key result of our approach as it shows the performance improvement obtained through face valuing.}
%\label{fig:infiniteobjects}
%\end{figure*}

\begin{figure}[t]
%\centering
\includegraphics[keepaspectratio,width=7.5cm,height=7.5cm]{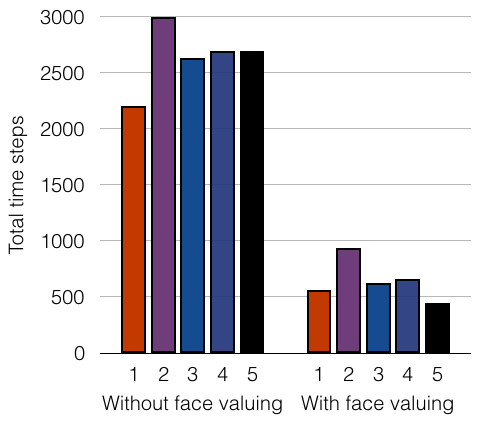}
\caption{Experiment with Infinite objects \& 5 grips: Plot of total time steps for limited grips and infinite objects setting. An experiment setting where a new object is introduced every episode and the agent needs to grasp this object from one of its grip. Each bar represents the total time steps taken by the agent to complete 15 episodes of this task. The plot was generated from data obtained from 5 independent runs. This is a key result of our approach as it clearly shows the performance improvement obtained through face valuing.}
\label{fig:infiniteobjects}
\end{figure}

%\begin{figure}[h]
%\centering
%\includegraphics[keepaspectratio,width=5cm,height=5cm]{figures_upd/infobj_learningcurve.eps}
%\caption{Infinite objects \& 5 grips experiment learning curves:Plot of total time steps for limited grips and infinite objects setting: an experiment setting where a new object is introduced every episode and the agent needs to grasp this object from one of its grip. Each bar represents the total time steps taken by the agent to complete 15 episodes of this task. The plot was generated from data obtained from 5 independent runs. This is a key result of our approach as it shows the performance improvement obtained through face valuing.}
%\label{fig:infiniteobjects_learningcurve}
%\end{figure}

A second experiment (Fig. \ref{fig:infiniteobjects}) showed the performance improvement obtained through our approach in a different and more difficult setting: one where a new object was generated for each episode and no object was ever seen more than once by the agent throughout the experiment. This experiment therefore explored the ability of face valuing to address new or changed tasks, and highlighted the importance of adapting quickly to a user\rq s preferences.

The Fig. \ref{fig:infiniteobjects} denotes the total time steps taken by the agents to complete this experiment whereas the Table \ref{tab:totalrewardsinfobj} shows the total number of instances of human explicit feedback required by the agent to successfully complete this task. Data was generated from experiments with a single user.

From the plot in Fig. \ref{fig:infiniteobjects}, it can be observed that the face-valuing agent was much quicker in adapting to the user\rq s preferences. It learned to complete the task quicker than an agent without face valuing. From the Table \ref{tab:totalrewardsinfobj}, it is clear that that total number of instances of human generated feedback to complete this task was less for the face-valuing agent. 

Since a new object was introduced in every episode, the agent without face valuing could not learn the possible grip/object combinations only from human-generated rewards. This was the cause for it requiring more  human feedback in completing this task. In the face-valuing approach, as the learning agent relied on values related to facial features, it could adapt easily in these situations. Effectively, the agent with face valuing learned to keep switching the grips periodically until the user gave a ``go ahead" expression. Unfortunately, the agent without face valuing did not have this advantage and could not perform effectively in this setting. 

\begin{table}[h!]
\centering
\begin{adjustbox}{width=0.4\textwidth}
\small
 \begin{tabular}{| c c |} 
 \hline
  Agent & Total no. of button pushes \\
 \hline \hline
w/o face valuing & 623.6 \\
w/ face valuing & 137.4 \\
 \hline
 \end{tabular}
 \end{adjustbox}
 \caption{Total number of button pushes provided for adapting an agent to the user\rq s preference in experiment 2. The values represent the total number of button pushes given by the human user to an agent for the complete experiment setting that lasted 15 episodes. Average obtained from 5 independent runs.}
 \label{tab:totalrewardsinfobj}
\end{table}

\section{Discussion}

In our experiments, the learning agent with face valuing had the ability to perceive a human user\rq s face and, eventually, learned to perceive a value of its behavior from its user's facial expressions. Our preliminary results therefore suggest that, by learning to value a human user\rq s facial expressions, the agent could adapt quickly to its user\rq s preferences with minimal explicit corrective feedback. This learning occurred as follows: during the initial phase of the experiment, the agent used the explicit corrective feedback to learn a value function from the user\rq s facial gestures; these gestures served as useful clues about future rewards based on the agent\rq s current behavior, and guided it\rq s behavior.

Several studies have shown that users, to a certain extent, are willing to teach machines to perform tasks automatically. For example, in medical domains, it is already common for people with amputations to extend their capabilities or limitations through a partnership with machines (Williams, T. W., 2011). However, currently available technology does not identify and adapt quickly to the different preferences of their users; this is a serious bottleneck to intelligence or physical amplification in human-machine partnerships. Our work helps begin to address these limitations.

%Specifically in the experiments presented above, we explored one specific idea wherein an agent learning to adapt to user\rq s preferences by perceiving a value of its actions from the user\rq s facial expressions. However, any other source that effectively reflects a user\rq s intents and preferences could also be used in the same manner. 

%In short, the task requires the agent to select the user\rq s preferred grip for a given object, then move towards the object and grasp it. The task is nontrivial because, potentially there can be many correct grip-object combinations and the agent has to figure out the right grip for the given object by interacting with its human user. Specifically, 

For evaluating our approach, we introduced a grip selection task wherein the learning agent had to figure out the goal through its interaction with the user; this agent can be readily termed a {\em goal-seeking agent} (Pilarski et al., 2015). To demonstrate the significance of our approach, we performed two sets of experiments: the first one involved multiple object-grip settings on the grip selection task; we termed the second experiment as the {\em infinite objects setting}, because one new object was generated for every episode and the agent needed to grasp this object by selecting one grip from its limited set of grips. This infinite objects settings is pertinent to real-world scenarios, where there are uncountable number of objects which can be grasped from a limited set of grip patterns. %There could also be multiple grip patterns that are suitable towards grasping a given object.

The results from the first user experiment (Fig. \ref{fig:learningcurves} (a), (b), $ \cdots $, (i)) suggest that the face-valuing agent learns to adapt quicker to its user\rq s preferences in this task. From the plots in Fig. \ref{fig:learningcurves} (j), (k), (l), it can be observed that the face-valuing agent learns to adapt to its user\rq s preferences with significantly lesser number of explicit human generated feedback signals, particularly in difficult experiment settings. From the second user experiment (Fig. \ref{fig:infiniteobjects}), we empirically show a scenario where conventional methods can fail. From both the experiments, it can be observed that the face-valuing agent successfully learns to adapt and completes one episode after another by relying only on facial expressions, specifically the value learned from facial expressions. On the other hand, the agent without face valuing could rely only on human generated feedback for identifying the correct grip for a given object, which resulted in a greater number of button pushes being given by the user. Moreover, we observed that the face valuing agent learned to wait for an affirmative facial expression before moving towards the object. Otherwise, the agent would switch from one grip to another at the grip-changing station until the user provided an affirmative expression. 

%While the present approach learned a state-action value function of a user\rq s facial expressions,  been improved to learn a general state-value function. Additionally, by using better feature representations, our approach can be extended to recognize subtle facial expressions made by humans.

Though our experiments were simulated, we believe that our approach can be much more valuable in a realistic robot setting---we expect a robot\rq s behavior would elicit more expressive facial feedback from the user than our simple simulated domain, and thus more powerful features for a face-valuing agent. In a robot setting, the user can observe the robot\rq s actions and their consequences in a real-world environment, where it is  natural for the user to implicitly emote through various facial cues. Robotic experiments are needed to help quantify the advantage of a face-valuing approach to human-machine interaction. 

%
%There has been many studies in this domain, but most of them rely on predefined vocabulary or push buttons alone for interacting with a learning agent.  We have shown that these conventional approaches fail in difficult tasks mimicking real-world scenarios. 

\section{Conclusions}

We introduced a new and a promising approach called {\em face valuing} for adapting an agent to a user\rq s preferences, and showed that it can produce large performance improvements over a conventional agent that learns only from human-generated rewards. By allowing the agent to perceive a value from a user\rq s facial expressions, the total number of expensive human generated-rewards delivered during a task was substantially reduced and the agent was quickly able to adapt to its user's preferences. Face valuing learns a mapping from facial expression to user satisfaction; it formalizes satisfaction as a value function and learns this value function through temporal-difference methods. Most work on the use of facial features in human-machine interaction uses facial features as control signals for an agent; surprisingly, our work seems to be the first to use facial expressions to instead train a learning system. Face valuing is general and largely task agnostic, and we believe it will therefore extend well to other settings and other forms of human body language.

\clearpage

%\bibliographystyle{named}
%\bibliography{ijcai16}

%\bibitem{} Mohan, S., Mininger, A., Kirk, J. and Laird, J.E., 2012, October. Learning Grounded Language through Situated Interactive Instruction. In AAAI Fall Symposium: Robots Learning Interactively from Human Teachers.

%\bibliographystyle{authordate1}

\end{document}